\documentclass[aps,prl,reprint,superscriptaddress,letterpaper,twocolumn,longbibliography,floatfix]{revtex4-2}

\usepackage{polyglossia}
\setmainlanguage{english}

\usepackage{graphicx}
\usepackage[svgnames,dvipsnames,x11names]{xcolor}
\usepackage[colorlinks=true,linkcolor=SpringGreen4,citecolor=blue,urlcolor=Magenta3]{hyperref}

\usepackage{amsmath}
\usepackage{amssymb}
\usepackage{float}
\usepackage{amsthm}
\usepackage{amsfonts}
\usepackage{mathtools}
\usepackage{mathrsfs}
\usepackage{bm}
\usepackage{bbm}
\usepackage{empheq}
\usepackage{cases}
\usepackage{euscript}

\usepackage[draft,todonotes={textsize=scriptsize}]{changes}
\setuptodonotes{inline}
\definechangesauthor[name={Razvan Stanescu}, color=Blue4]{RS}
\definechangesauthor[name={Hugo Ribeiro}, color=SpringGreen4]{HR}

\newcommand{\bra}[1]{\langle #1|}
\newcommand{\ket}[1]{|#1\rangle}

\newcommand{\ketbra}[2]{\ket{#1}\!\bra{#2}}
\newcommand{\norm}[1]{\left\lVert#1\right\rVert}
\newcommand{\mm}[1]{\mathrm{#1}}
\newcommand{\abs}[1]{\left|#1\right|}

\newcommand{\di}[1]{\mathop{}\!\mathrm{d} #1}

\DeclareMathOperator{\Tr}{Tr}
\DeclareMathOperator{\BigO}{\mathcal{O}}

\def \ud{\mathrm{d}}

\def \uf{\mathrm{f}}

\def \uF{\mathrm{F}}

\def \uI{\mathrm{I}}

\def \cv{\mbox{\boldmath$c$}}

\def \hOmega{\hat{\Omega}}

\def \hn{\hat{n}}
\def \hA{\hat{A}}

\def \hE{\hat{E}}
\def \hH{\hat{H}}

\def \hO{\hat{O}}
\def \hP{\hat{P}}
\def \hQ{\hat{Q}}

\def \hT{\hat{T}}
\def \hU{\hat{U}}
\def \hV{\hat{V}}
\def \hW{\hat{W}}

\begin{document}

\title{Error-Generator-Level Compression for Fast Quantum Gates}

\author{Razvan Stanescu}
\affiliation{Department of Physics and Applied Physics, University of Massachusetts Lowell, Lowell, MA 01854, USA}
    
\author{Hugo Ribeiro}
\affiliation{Department of Physics and Applied Physics, University of Massachusetts Lowell, Lowell, MA 01854, USA}

\begin{abstract}
Fast quantum gates in multilevel systems require suppressing many coherent-error channels using pulses that remain simple to
implement. We introduce error-generator-level compression, a Magnus-based control principle that minimizes the complete projected
error generator at a chosen perturbative order while restricting the implemented waveform to a few independently adjustable
coefficients. This construction preserves all modeled computational and leakage errors rather than truncating the pulse by error
channel, while its generator-level objective directly penalizes residual eigenphases that can signal departure from the
perturbative regime. For fast transmon gates, a three-parameter pulse suppresses errors by orders of magnitude, outperforms
grid-optimized leading-order DRAG over the gate-time range studied, and approaches a fully parameterized 17-parameter Magnus
correction. The compressed pulse is also substantially less sensitive to the illustrative finite-bandwidth filtering model
considered here.
\end{abstract}

\maketitle

\paragraph*{Introduction.---}
Fast quantum control in multilevel systems is limited by coherent error channels absent from an ideal qubit description. In weakly
anharmonic superconducting circuits, microwave drives intended to address the computational transition can also induce leakage,
unwanted phases, and off-resonant couplings to noncomputational states. Analytic corrections such as DRAG, experimentally
optimized two-quadrature pulses, and recent frequency-shaped extensions suppress leakage using simple, experimentally natural
envelopes~\cite{Motzoi2009PRL,Gambetta2011PRA,Chow2010PRA,Theis2018EPL,Hyyppa2024PRXQ}. These methods highlight an essential
requirement for scalable control:~High-fidelity pulses must remain simple enough to calibrate and implement.

Quantum optimal-control methods instead optimize pulse shapes against a final-time state-transfer or gate fidelity. Gradient,
reduced-basis, and analytic-parameter methods have become central tools for quantum
technologies~\cite{Khaneja2005JMR,Krotov1996Book,Caneva2011PRA, Machnes2018PRL,Glaser2015EPJD,Muller2022RPP}, producing fast,
low-leakage gates with experimentally constrained parameterizations~\cite{Werninghaus2021NPJ,Song2022PRA}. Machine-learning
characterization can further reduce model error~\cite{Genois2025PRApplied}. These approaches generally optimize the realized
unitary rather than the perturbative error generator, which can obscure the physical origin of the correction and require
substantial calibration.

\begin{figure}[t]
    \centering
    \includegraphics[width=0.99\columnwidth]{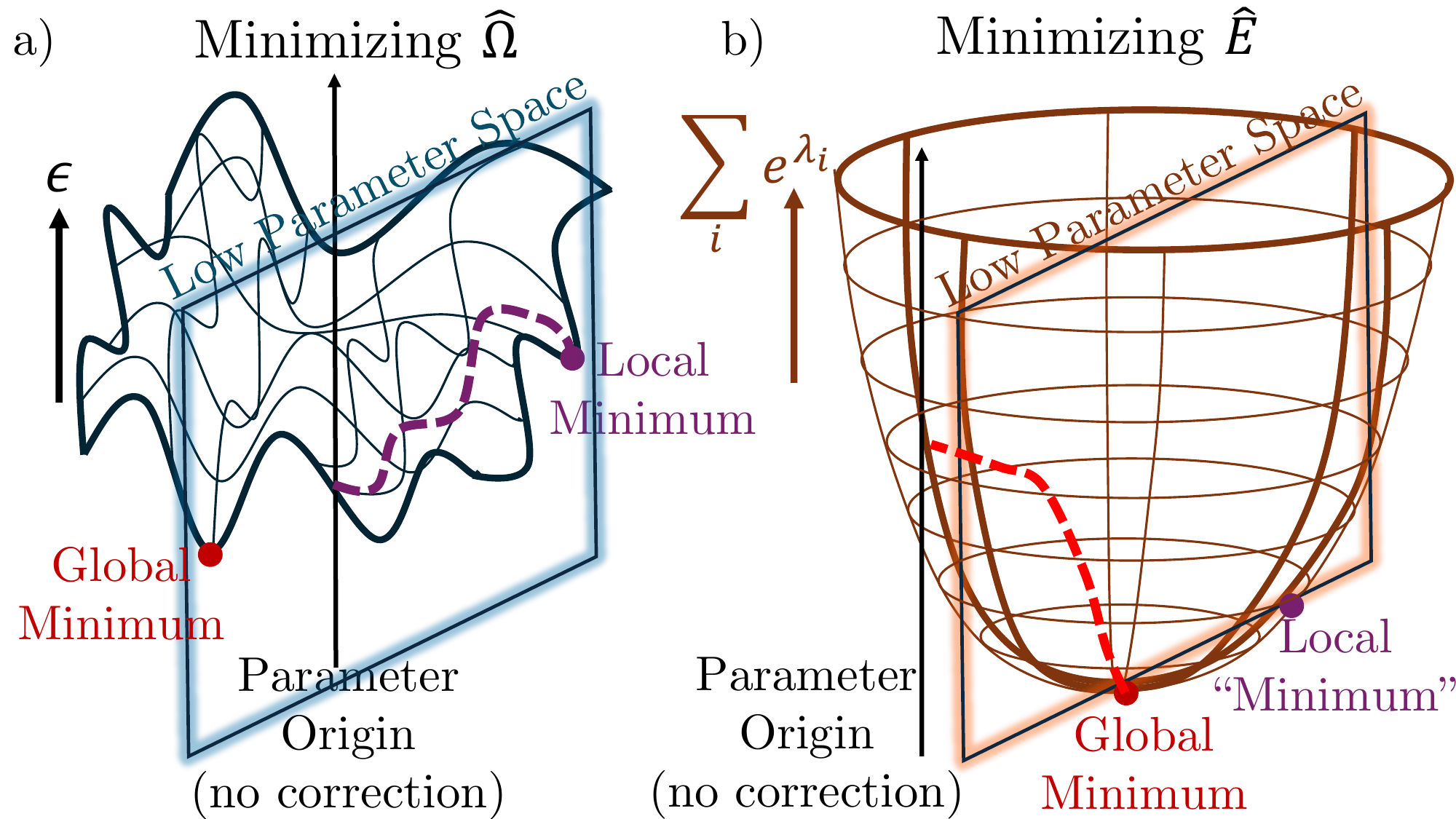}
    \caption{
        Schematic comparison between fidelity-based compression and generator-level compression. Minimizing a final-time unitary
        objective can favor nonperturbative solutions because the unitary is periodic in the residual eigenphases. In contrast,
        compressed Magnus control evaluates a spectral penalty that is convex in the Hermitian diagnostic generator and is
        minimized when the projected residual eigenphases vanish; the induced landscape in pulse coefficients need not be convex.
    }
    \label{fig:Concept}
\end{figure}

Pulse simplicity is also experimentally consequential. Room-temperature electronics, cryogenic wiring, and on-chip transfer
functions distort control waveforms, making accurate delivery a nontrivial calibration problem~\cite{Rol2020APL,Guo2024PRApplied}.
Strong drives can additionally activate parasitic, multiphoton, and inelastic transition pathways~\cite{Dai2026PRX}. The central
challenge is therefore to suppress a high-dimensional error model using a low-dimensional waveform.

Magnus-based control addresses this problem by modifying the residual interaction-picture dynamics. Previous constructions
suppress spectrally crowded transitions, leakage, and nonadiabatic errors~\cite{Schutjens2013PRA,Ribeiro2017PRX}, while related
perturbative and geometric approaches cancel accumulated error operators~\cite{Buterakos2021PRXQ,Watanabe2024PRA}.
Constrained-control methods generate unavailable directions through higher-order commutators~\cite{Roque2021NPJ}, but fully
parameterized solutions can still require many waveform components.

A naive approach is to retain only the largest components of a Magnus correction, but this is uncontrolled. Higher-order
commutators can regenerate canceled operators and couple nominally subleading channels into dominant errors. Optimizing the
remaining coefficients against gate fidelity avoids this truncation but abandons the generator-level perturbative structure.
Moreover, because unitary fidelity is periodic in the residual eigenphases, it can favor large generators outside the perturbative
regime.

The main contribution of this Letter is to compress the implemented waveform without compressing the perturbative error model. We
restrict the pulse to a few independently adjustable coefficients while retaining every computational-subspace error and leakage
coupling in the projected generator at the chosen Magnus order. The coefficients minimize a spectral penalty that is convex in the
Hermitian diagnostic generator, though not necessarily in coefficient space. Near the perturbative origin, this penalty reduces to
the squared Frobenius norm of the projected generator, which is directly related to the leading average fidelity error. Away from
the origin, the penalty increasingly suppresses large residual eigenphases.

We demonstrate the method for fast transmon gates. Three-parameter compressed pulses suppress errors by orders of magnitude,
outperform naive truncation and grid-optimized leading-order DRAG over the gate-time range studied, approach the performance of a
fully parameterized $17$-parameter fourth-order Magnus correction, and are substantially less sensitive to the finite-bandwidth
filtering model considered here.

\paragraph*{Compressed generator minimization.---}
We build on the constrained Magnus-control framework of Refs.~\cite{Ribeiro2017PRX,Roque2021NPJ}.  We write
\begin{equation} 
    \hH (t) = \hH_0 (t) + \varepsilon \hV (t) + \sum_{k \geq 1} \varepsilon^k \hW_k (t), 
    \label{eq:Hsplit}
\end{equation}
where $\hH_0 (t)$ generates the target operation at $t=t_\uf$, $\hV (t)$ contains the coherent error terms, and $\hW_k (t)$ are
control Hamiltonians restricted to the experimentally available controls. The parameter $\varepsilon$ is a formal bookkeeping
parameter, set to one at the end, which marks $\hV (t)$ and the control Hamiltonians $\hW_k (t)$ as small compared with the
target dynamics generated by $\hH_0 (t)$.

The role of the control Hamiltonians is to average away the effect of $\hV (t)$ over the full evolution without modifying the
desired operation generated by $\hH_0 (t)$. To express this condition, we first consider the full series for the control
Hamiltonian. Moving to the interaction picture generated by $\hH_0 (t)$ gives $\hH_\uI (t) = \varepsilon \hV_\uI (t)
+ \sum_{k\geq 1} \varepsilon^k \hW_{k,\uI}(t)$. Exact cancellation would require the corresponding final-time evolution to satisfy
\begin{equation}
    \hU_\uI (t_\uf) = \hT \exp\left[ -i \int_0^{t_\uf} \di{t} \hH_\uI (t)  \right] = \mathbbm{1},
    \label{eq:UIidentity}
\end{equation}
up to irrelevant phases. Because enforcing this condition exactly is generally as difficult as solving the driven problem, Magnus
control enforces it perturbatively, order by order. At order $n$, one uses
\begin{equation}
    \hH_\uI^{(n)}(t) = \varepsilon \hV_\uI (t) + \sum_{k=1}^{n} \varepsilon^k \hW_{k,\uI} (t),
    \label{eq:HIn}
\end{equation}
and writes the corresponding truncated Magnus generator as 
\begin{equation}
    \hOmega^{(n)} (t_\uf) = \sum_{k=1}^{n} \varepsilon^k \hOmega^{(n)}_k (t_\uf).
    \label{eq:MagnusTrunc}
\end{equation}
The condition
\begin{equation}
    \exp\left[\hOmega^{(n)} (t_\uf)\right] = \mathbbm{1} + \mathcal{O}\left(\varepsilon^{n+1}\right)
    \label{eq:MagnusIdentity}
\end{equation}
is imposed recursively:~After $\hW_1 (t), \ldots, \hW_{n-1} (t)$ have been fixed, $\hW_n (t)$ is chosen so that the relevant
order-$\varepsilon^n$ terms in the truncated Magnus generator vanish.

The operator cancellation conditions are converted into equations for scalar coefficients by decomposing both the error
Hamiltonian and the allowed correction Hamiltonian in the same operator basis. In the lab frame, let
$\{\hA_\mu\}_{\mu \in B}$ denote a convenient operator basis.  We write 
\begin{equation}
    \begin{aligned}
        \hV (t) &= \sum_{\mu \in E} v_\mu (t) \hA_\mu, \\
        \hW_n (t) &= \sum_{\mu \in C} w_\mu^{(n)} (t) \hA_\mu .
    \end{aligned}
    \label{eq:lab_decomposition}
\end{equation}
Here, $E \subseteq  B$ denotes the set of error channels present in $\hV (t)$, while $C \subseteq B$ denotes the subset of
operators that couple to external control fields. The fact that $C$ may not span all projected error directions is the origin of
the constrained-control problem. Transforming Eq.~\eqref{eq:lab_decomposition} to the interaction picture gives the corresponding
decomposition of $\hV_\uI (t)$ and $\hW_{n,\uI} (t)$.

Following Ref.~\cite{Roque2021NPJ}, the unknown correction fields are expanded in a Fourier basis,
\begin{equation}
        w_\mu^{(n)} (t) = \sum_{m=1}^{M_\mu} a_{\mu, m}^{(n)} \left[ 1 - \cos\left( \omega_m t\right)\right] 
        + \sum_{l=1}^{L_\mu} b_{\mu,l}^{(n)} \sin\left(\omega_l t \right),
    \label{eq:FourierCorrection}
\end{equation}
with $\mu\in C$ and $\omega_j = 2\pi j /t_\uf$. Thus, the condition $\hOmega^{(n)} (t_\uf) = \mathbf{0}$, projected onto the error
components associated with the target subspace, becomes a finite set of algebraic equations for the Fourier coefficients
$\{a_{\mu,m}^{(n)},\, b_{\mu,l}^{(n)}\}$. Note that the Fourier decomposition of $w_\mu^{(n)} (t)$ ensures that the control fields
vanish at the beginning and end of the protocol, $w_\mu^{(n)} (0) = w_\mu^{(n)} (t_\uf) = 0$.

If the available controls span the projected order-$\varepsilon^n$ error directions in the interaction picture, the cancellation
equations are linear in the Fourier coefficients. Otherwise, the problem is singular or ill conditioned. Missing error operators
must be generated through higher-order Magnus commutators, leading to nonlinear algebraic equations~~\cite{Roque2021NPJ}.

Fully parameterized Magnus corrections can require many Fourier coefficients and are consequently harder to synthesize and more
sensitive to waveform distortion. We instead vary only a small subset of coefficients while evaluating the complete projected
truncated generator analytically. The error model therefore remains high-dimensional while the implemented waveform is
low-dimensional.

Since $\hOmega^{(n)} (t_\uf)$ is anti-Hermitian, we define the Hermitian truncated error generator
\begin{equation}
    \hE^{(n)}(t_\uf) = -i \hOmega^{(n)} (t_\uf).
    \label{eq:Edef}
\end{equation}
This diagnostic generator is not an additional physical Hamiltonian; rather, it compactly describes the coherent error accumulated
over the gate, with real eigenvalues equal to the residual eigenphases. A small $\hE^{(n)}$ ensures that the residual evolution
remains near the perturbative identity solution. By contrast, the residual unitary can appear close to the identity even for a
large generator because its eigenphases are periodic, for example near $2\pi$. We therefore construct the compressed objective
from $\hE^{(n)}$, rather than from the residual unitary alone, and next isolate the components that affect the target operation.

Let $\hP$ project onto the subspace of interest and $\hQ=\mathbbm{1}-\hP$ onto its complement. The $\hP\hE^{(n)}\hP$ block
describes coherent errors within the computational subspace, while the $\hP\hE^{(n)}\hQ$ and $\hQ\hE^{(n)}\hP$ blocks describe
leakage couplings. We retain these blocks while discarding dynamics entirely internal to $\hQ$, and remove the component
proportional to $\hP$ because it produces only a common phase. This defines
\begin{equation}
    \hE_{\mm{rel}}^{(n)} = \mathcal{P}_\mm{rel}\left[\hE^{(n)}\right],
    \label{eq:Erel}
\end{equation}
where $\mathcal{P}_\mm{rel}[\hO] = \hP \hO \hP+ \hP \hO \hQ + \hQ \hO \hP - [\Tr(\hP \hO \hP)/\Tr (\hP)] \hP$.

We collect the retained Fourier coefficients at order $n$ into the vector $\cv^{(n)} = \{a_{\mu,m}^{(n)},\,b_{\mu,l}^{(n)}\}_{\mu
\in C}$. The compressed coefficients minimize the generator-level cost 
\begin{equation}
    \cv_\star^{(n)} = \operatorname*{arg\,min}_{\cv^{(n)}} \Phi^{(n)}\left[\cv^{(n)}\right],
    \label{eq:argmin}
\end{equation}
with
\begin{equation}
    \Phi^{(n)}\left[\cv^{(n)}\right] = \Tr\left( e^{\hE_{\mm{rel}}^{(n)} \left[\cv^{(n)}\right]} +
    e^{-\hE_{\mm{rel}}^{(n)}\left[\cv^{(n)}\right]} - 2\mathbbm{1}_{\mm{supp}} \right).
    \label{eq:Phi}
\end{equation}
Here, $\mathbbm{1}_{\mm{supp}}$ is the identity on the support of $\hE_{\mm{rel}}^{(n)}$. If $\lambda_j$ are the eigenvalues of
$\hE_{\mm{rel}}^{(n)}$, then
\begin{equation}
    \Phi^{(n)} = 2\sum_j \left( \cosh\left[ \lambda_j^{(n)}\right] - 1 \right).
    \label{eq:PhiEig}
\end{equation}
Thus, $\Phi^{(n)}$ is a nonnegative spectral penalty minimized when the projected residual eigenphases vanish. The method does not
make the truncated Magnus expansion exact or remove terms beyond the retained order. Rather, it selects a low-dimensional pulse by
suppressing all components of the projected truncated diagnostic generator while penalizing nonperturbatively large residual
eigenphases. As derived in the Supplemental Material~\cite{Supplemental}, expanding the cost near the perturbative origin gives
\begin{equation}
    \begin{aligned}
        \Phi^{(n)} &= \Tr\left[ \left( \hE_{\mm{rel}}^{(n)} \right)^2 \right] + \BigO\left[\left(\hE_{\mm{rel}}^{(n)} \right)^4
        \right] \\
        &=\norm{\hE_{\mm{rel}}^{(n)}}_\uF^2 + \BigO \left( \norm{\hE_{\mm{rel}}^{(n)}}_\uF^4 \right).
    \end{aligned}
    \label{eq:Frobenius}
\end{equation}
Thus, locally, $\Phi^{(n)}$ penalizes the combined magnitude of computational-subspace errors and leakage couplings through the
squared Frobenius norm of the projected diagnostic generator. For larger residual eigenphases, the hyperbolic-cosine form grows
more rapidly and discourages solutions outside the perturbative regime. The objective is convex in the diagnostic generator and
uniquely minimized when the projected generator vanishes, although its dependence on the pulse coefficients need not be globally
convex.

The functional $\Phi^{(n)}$ is convex with respect to the Hermitian diagnostic generator. Convexity in the Fourier coefficients
follows only when $\hE_{\mm{rel}}^{(n)}$ depends affinely on them; in singular or ill-conditioned cases, higher-order commutators
generally generate polynomial dependence (see Supplemental Material~\cite{Supplemental}). 

\paragraph*{Transmon model and pulse parameterization.---}
As an example application, we consider a single-transmon gate because it provides a direct route to experimental implementation
while retaining the central constrained-control problem:~A limited set of physical controls must suppress many coherent-error
channels. This setting provides a transparent test of generator-level compression without restricting the method to single-qubit
errors. The compression principle itself is not transmon-specific and can incorporate additional coherent-error channels,
including crosstalk, whenever they can be represented in a truncated Magnus generator with a restricted control parametrization.

Moderately fast transmon gates require a multilevel description. We retain the four lowest eigenstates, comprising the
computational subspace and two leakage levels, because a three-level truncation is not quantitatively accurate in the regime
studied here~\cite{Stanescu2026}. The resulting projected generator contains $11$ independent error components, some of which
involve the second leakage level and arise only through higher-order Magnus commutators. These components define $11$ coupled
nonlinear cancellation conditions, not a one-to-one parametrization of the waveform. Each waveform coefficient generally affects
several error components, while indirectly generated components can depend on multiple pathways. We therefore use $17$
independently adjustable parameters for the fully parameterized fourth-order reference correction; details are given in the
Supplemental Material~\cite{Supplemental}.

We work in the eigenbasis of the transmon model~\cite{koch2007,blais2021,makhlin2001} for $E_J/E_C = 50$. We set $\hbar=1$, so
energies and angular frequencies are expressed in the same units. In the four-level truncation, the driven Hamiltonian is
\begin{equation}
    \hH(t) = \sum_{i=0}^3 \omega_i \ketbra{i}{i} + f(t) \sum_{i,j=0}^3 n_{i,j} \ketbra{i}{j},
    \label{eq:transmon_drive}
\end{equation}
where $\omega_i$ are the transmon eigenfrequencies and $n_{ij} = \bra{i}\hn\ket{j}$ are the charge-operator matrix elements
evaluated using the transmon eigenstates $\ket{i}$ and $\ket{j}$. We define $\omega_{ij}=\omega_j-\omega_i$. Because the
four-level model has more than one anharmonicity, we use $\alpha_2 = \omega_{12} - \omega_{01}$, the anharmonicity associated with
the first leakage level $\ket{2}$, to report the dimensionless gate time $\abs{\alpha_2} t_\uf$. This directly measurable quantity
sets the dominant leakage scale for the gates considered below. The retained eigenfrequencies and charge-matrix elements are
listed in the Supplemental Material~\cite{Supplemental}.

The microwave drive is written in terms of its two quadratures, $f (t) = f_x (t) \cos(\omega_\ud t) + f_y (t) \sin(\omega_\ud t)$. 
We choose the uncorrected envelopes to be
\begin{equation}
    \begin{aligned}
        f_x (t) &= \frac{\theta_0}{n_{0,1} t_\uf} \left[ 1 - \cos\left( \frac{2 \pi t}{t_\uf} \right)  \right],\\
        f_y (t) &= 0,
    \end{aligned}
    \label{eq:OriginalEnv}
\end{equation}
where $\theta_0 = \pi/2$ is the target rotation angle about the $x$ axis of the Bloch sphere.

Because the correction is applied through the same physical charge-drive channel, it can modify only the two microwave
quadratures,
\begin{equation}
    \hW (t) = \left[ g_x (t) \cos(\omega_\ud t) + g_y (t) \sin(\omega_\ud t)\right] \hn.
    \label{eq:transmon_correction}
\end{equation}
The modified quadrature envelopes are therefore $f_{x,\mm{mod}} (t) = f_x (t) + g_x (t)$ and $f_{y,\mm{mod}} (t) = f_y (t) + g_y
(t)$. The correction envelopes are expanded in the Fourier basis of Eq.~\eqref{eq:FourierCorrection}, and the retained
coefficients enter the compressed parameter vector $\cv^{(n)}$. We keep the drive frequency constant throughout the pulse and
write $\omega_\ud = \omega_{01} + \Delta$. The constant detuning $\Delta$ is therefore an additional control parameter, avoiding the
experimental complexity of a chirped carrier.

For the compressed pulse used in the main comparison, we retain only three free parameters. The two correction envelopes are each
parametrized by a single Fourier coefficient, $g_x(t) = a_x [1 - \cos(2\pi t/t_\uf)]$ and $g_y (t) = b_y \sin(2\pi t/t_\uf)$,
together with the constant detuning $\Delta$. This choice is deliberately minimal:~$\Delta$ adds no time-dependent spectral
content, the $g_x (t)$ correction has the same fundamental frequency as the uncorrected envelope $f_x (t)$, and the $g_y (t)$
correction contains only the lowest sine harmonic. The three-parameter pulse therefore tests whether all projected error
components at the retained Magnus order can be suppressed using a waveform with minimal experimentally natural spectral content.

For comparison, we use leading-order DRAG as the practical low-complexity benchmark, since it remains a standard correction
calibrated in superconducting-qubit experiments ~\cite{Motzoi2009PRL,Chow2010PRA,Chen2016PRL},
\begin{equation}
    g_{\mm{DRAG},y} (t) = -\frac{\beta}{\alpha_2} \dot{f}_x (t),
    \label{eq:drag}
\end{equation}
where $\beta$ is a dimensionless DRAG coefficient. At each gate time, we perform a grid search over $\beta=0,0.1,\ldots,1.0$ and
retain the value yielding the smallest average fidelity error. This procedure allows the preferred DRAG coefficient to vary with
gate duration without invoking continuous optimization. The complete coefficient scan is given in the Supplemental
Material~\cite{Supplemental}.

We quantify gate performance using the average fidelity error $\epsilon=1- (\Tr(\hO\hO^\dagger)+|\Tr(\hO)|^2)/[d(d+1)]$, where
$\hO=\hU_{\mm{ideal}}^\dagger\hP\hU(t_\uf)\hP$, $\hU_{\mm{ideal}}$ is the target operation, and $\hP$ projects onto the $d=2$
computational subspace. This metric includes both coherent errors within the computational subspace and population lost to
leakage. In the perturbative regime, minimizing $\Phi^{(n)}$ minimizes the squared Frobenius norm of the projected diagnostic
generator and hence the combined magnitude of these errors. At the optimum, their leading quadratic contributions satisfy
$\Phi^{(n)}(\cv_\star^{(n)})/4 \lesssim\epsilon\lesssim \Phi^{(n)}(\cv_\star^{(n)})/3$. Thus, reducing the generator-level cost
reduces the average fidelity error in the perturbative regime; the derivation is given in the Supplemental
Material~\cite{Supplemental}.

\begin{figure}[t]
    \includegraphics[width=0.99\columnwidth]{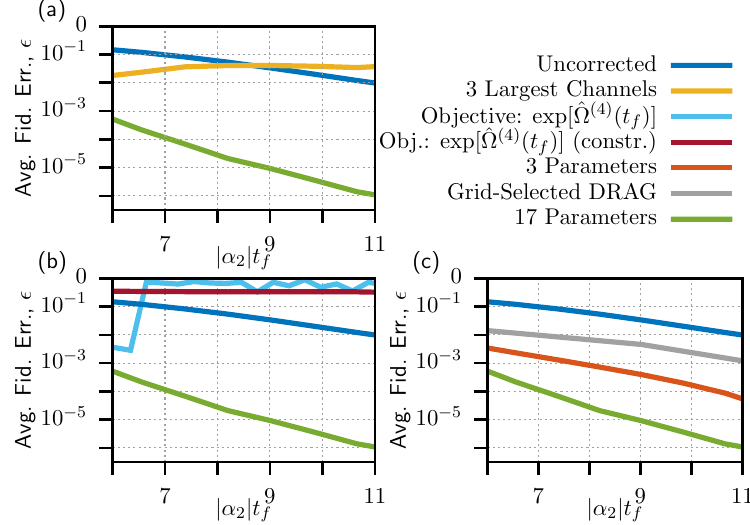}
    \caption{
        Average fidelity error $\epsilon$ versus dimensionless gate time $\abs{\alpha_2}t_\uf$ for a target $\pi/2$ rotation about
        the $x$ axis, computed from the projected final-time evolution of the four-level transmon model. (a) Naive three-parameter
        truncation retaining only the largest error channels predicted by the fourth-order generator (yellow). (b) Three-parameter
        minimization of the truncated residual unitary $\exp[\hOmega^{(4)}]$ without (light blue) and with (red) partial Magnus
        constraints. (c) Three-parameter compressed Magnus correction obtained by minimizing $\Phi^{(4)}$ (orange) and
        grid-optimized leading-order DRAG (gray), where the smallest error is selected independently at each gate time from the
        discrete scan $\beta=0,0.1,\ldots,1.0$. The uncorrected baseline (blue) and fully parameterized 17-parameter fourth-order
        Magnus correction (green) are repeated in all panels for reference.
    }
    \label{fig:performance}
\end{figure}

\paragraph*{Results.---}
Figure~\ref{fig:performance} compares reduced-parameter strategies evaluated using the fourth-order diagnostic generator and
associated objectives. The fully parameterized reference correction is obtained using the iterative singular construction of
Ref.~\cite{Roque2021NPJ}, as detailed in the Supplemental Material~\cite{Supplemental}. In Fig.~\ref{fig:performance}(a), this
17-parameter correction provides the reference performance (green trace), whereas a naive three-parameter truncation retaining
only the largest predicted error channels performs poorly (yellow trace) and can even exceed the uncorrected baseline at longer
gate times. This failure reflects the nonlinear structure of the Magnus expansion:~Suppressing selected channels does not minimize
all projected fourth-order error components, and higher-order commutators can regenerate errors.

Figure~\ref{fig:performance}(b) shows that minimizing the truncated residual unitary $\exp[\hOmega^{(4)}]$ is also insufficient
(light blue trace). At shorter gate times, the optimization can still locate the perturbative minimum near zero residual
eigenphase. At longer gate times, unitary periodicity and eigenphase cancellations create competing minima where the exponential
appears close to the identity despite a large diagnostic generator. Constraining the three largest predicted error components of
the fourth-order Magnus generator to vanish does not remove this failure mode (red trace), since the remaining components can
still generate large residual eigenphases masked by unitary periodicity.

By contrast, Fig.~\ref{fig:performance}(c) shows that minimizing $\Phi^{(4)}$ directly suppresses the eigenphases of
$\hE_{\mm{rel}}^{(4)}$ (orange trace). The DRAG benchmark (gray trace) is the pointwise lower envelope of the grid search
$\beta=0,0.1,\ldots,1.0$, with the best sampled coefficient selected independently at each gate time. At the shortest gate time
shown, $\abs{\alpha_2}t_\uf=5.74$, the average fidelity errors are $10^{-0.78}$ for the uncorrected pulse, $10^{-1.8}$ for the
grid-selected DRAG benchmark, $10^{-2.4}$ for the three-parameter compressed Magnus pulse, and $10^{-3.1}$ for the fully
parameterized 17-parameter Magnus correction. The compressed pulse therefore reduces the error by a factor of approximately $42$
relative to the uncorrected pulse and a factor of approximately $4$ relative to the grid-selected DRAG benchmark, while remaining
within a factor of approximately $5$ of the fully parameterized correction. At the longest gate times, the three- and 17-parameter
pulses differ by roughly two orders of magnitude, but this regime is less experimentally relevant because relaxation and dephasing
increasingly dominate the total error. The speed advantage is also visible at fixed error. An average fidelity error of $10^{-3}$
is reached at $\abs{\alpha_2} t_\uf \simeq 7.8$ with the three-parameter compressed pulse, compared with $\abs{\alpha_2}t_\uf
\simeq 11.5$ for the grid-selected DRAG benchmark. The compressed construction therefore achieves the same error with a gate time
approximately $32\%$ shorter.

\begin{figure}[t]
    \includegraphics[width=0.99\columnwidth]{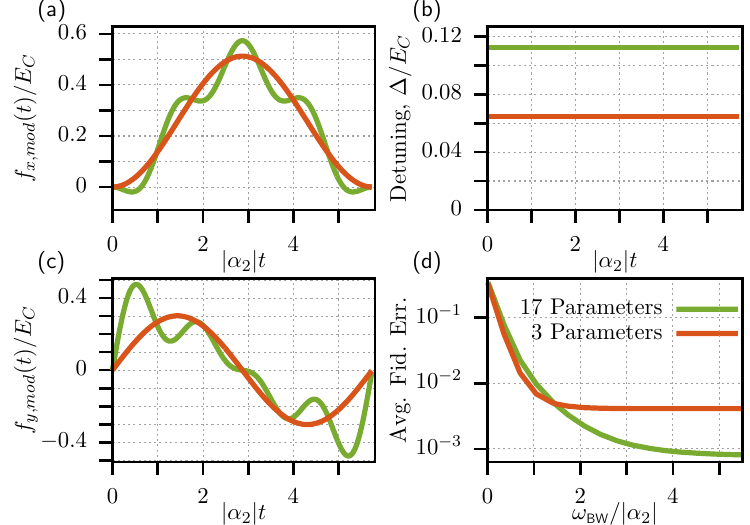}
    \caption{
        Modified pulses and finite-bandwidth performance at $\abs{\alpha_2} t_\uf=5.74$. The three-parameter compressed pulse is
        obtained with $a_x/E_C=-0.017$, $b_y/E_C=0.301$, and $\Delta/E_C=0.065$. Panels (a), (b), and (c) show the total modified
        in-phase envelope $f_{x,\mm{mod}}(t)$, constant detuning $\Delta$, and total modified quadrature envelope
        $f_{y,\mm{mod}}(t)$, respectively, for the fully parameterized 17-parameter Magnus pulse (green) and the three-parameter
        compressed Magnus pulse (orange). In panel (d), a Gaussian response with $3\,\mm{dB}$ bandwidth $\omega_{\mm{BW}}$ is
        applied to the complete modified baseband envelopes, while the carrier frequency and detuning remain unchanged. The
        filtered amplitudes and pulse areas are not renormalized, and the filter-induced Gaussian tails are truncated to the
        nominal gate interval; both effects are included in the average fidelity error computed from the filtered four-level
        evolution.
    }
    \label{fig:pulses}
\end{figure}

Figure~\ref{fig:pulses} compares the modified pulses at $\abs{\alpha_2}t_\uf=5.74$. The fully parameterized pulse (green trace)
contains visible higher-frequency structure across the control channels, whereas the three-parameter compressed pulse (orange
trace) is smooth and dominated by low-frequency components.

Figure~\ref{fig:pulses}(d) compares the sensitivity of these pulses under an illustrative finite-bandwidth filtering model:~A
Gaussian response is applied to the complete modified baseband envelopes $f_{x,\mm{mod}}(t)$ and $f_{y,\mm{mod}}(t)$, while the
carrier frequency and static detuning remain unchanged (see Supplemental Material~\cite{Supplemental}). The filtered amplitudes
and pulse areas are not renormalized, and the filter-induced tails are truncated to the nominal gate interval; these effects are
included in the reported average fidelity error. The fully parameterized pulse begins to degrade for
$\omega_{\mm{BW}}/\abs{\alpha_2}\lesssim5$, consistent with its fourth harmonic, $\omega_4/\abs{\alpha_2}=8\pi/5.74\simeq4.38$.
The compressed pulse degrades only near $\omega_{\mm{BW}}/\abs{\alpha_2}\sim1$, consistent with its lower-frequency content. This
provides a practical experimental advantage: the three-parameter pulse is substantially less sensitive to the finite-bandwidth
filtering model considered here.

\paragraph*{Discussion and conclusion.---}
Compressed Magnus control retains every projected error component through the chosen Magnus order without requiring an equally
high-dimensional waveform. The generator-convex spectral penalty suppresses residual eigenphases directly; coefficient-space
convexity is inherited only when the generator depends affinely on the pulse parameters.

For fast transmon gates, three-parameter pulses approach the fully parameterized fourth-order correction, outperform
grid-optimized leading-order DRAG over the gate-time range studied, and are substantially less sensitive to the finite-bandwidth
filtering model considered here. The method therefore combines high-dimensional error models with low-dimensional, experimentally
natural waveforms expected to reduce calibration complexity.

\bibliography{CompressedMagnus.bib}

\clearpage
\onecolumngrid
\begin{center}
    \textbf{\large Supplemental Material for:~``Error-Generator-Level Compression for Fast Quantum Gates''}
\end{center}

\setcounter{secnumdepth}{2}
\setcounter{equation}{0}
\setcounter{figure}{0}
\renewcommand{\theequation}{S\arabic{equation}}
\renewcommand{\thefigure}{S\arabic{figure}}

\section{Transmon parameters used in the numerical simulations} 
\label{app:transmon_parameters}

The numerical calculations use the four-level transmon model introduced in the main text. The transmon eigenstates,
eigenfrequencies, and charge-operator matrix elements are evaluated for $E_J/E_C=50$. We define the charging-energy angular
frequency $\omega_C$ through $E_C=\hbar\omega_C$ and set $\hbar=1$. Energies and angular frequencies are therefore expressed in
units of $\omega_C$. In particular, the dimensionless eigenfrequencies are $\omega_k/\omega_C$.

The charge-operator matrix elements are defined as
\begin{equation} 
    \hn = \sum_{i,j} n_{ij} \ketbra{i}{j}. 
    \label{eq:app_charge_operator_decomposition} 
\end{equation}
Because $\hn$ is Hermitian, the charge matrix elements satisfy $n_{ji}=n_{ij}^{*}$. For the real eigenstate convention used here,
$n_{ij}$ is real, and therefore $n_{ji}=n_{ij}$. Since $\hn$ is the dimensionless Cooper-pair-number operator, the matrix elements
$n_{ij}$ are dimensionless and are reported without an additional normalization by $\omega_C$. The numerical values used in the
simulations are listed in Table~\ref{tab:transmon_parameters}.

\begin{table}[b] 
    \begin{minipage}{0.58\linewidth} 
        \begin{ruledtabular} 
            \begin{tabular}{l l l} Quantity & Definition & Value \\ 
                \hline
                $\omega_0/\omega_C$ & Ground-state eigenfrequency & $-40.26$ \\ 
                $\omega_1/\omega_C$ & First excited-state eigenfrequency & $-21.31$ \\
                $\omega_2/\omega_C$ & Second excited-state eigenfrequency & $-3.52$ \\ 
                $\omega_3/\omega_C$ & Third excited-state eigenfrequency & $12.96$ \\
                \hline 
                $n_{01}$ & $\bra{0}\hn\ket{1}$ & $1.09$ \\ 
                $n_{02}$ & $\bra{0}\hn\ket{2}$ & $0$ \\ 
                $n_{03}$ & $\bra{0}\hn\ket{3}$ & $0.04$ \\ 
                $n_{12}$ & $\bra{1}\hn\ket{2}$ & $1.49$ \\ 
                $n_{13}$ & $\bra{1}\hn\ket{3}$ & $0$ \\ 
                $n_{23}$ & $\bra{2}\hn\ket{3}$ & $1.76$ 
            \end{tabular} 
        \end{ruledtabular}
    \end{minipage}
    \caption{
        Parameters of the four-level transmon model used in the numerical simulations for $E_J/E_C=50$. The charging-energy
        angular frequency is defined by $E_C=\hbar\omega_C$, with $\hbar=1$. Eigenfrequencies are reported in units of $\omega_C$
        and the charge-operator matrix elements $n_{ij}$ are dimensionless.
    }
    \label{tab:transmon_parameters} 
\end{table}

The transition frequencies and the anharmonicity associated with $\ket{2}$ are obtained from the eigenfrequencies in
Table~\ref{tab:transmon_parameters}:
\begin{equation} 
    \begin{aligned} 
        \omega_{01} &= \omega_1-\omega_0, \\ 
        \omega_{12} &= \omega_2-\omega_1, \\ 
        \omega_{23} &= \omega_3-\omega_2, \\ 
        \alpha_2 &= \omega_{12}-\omega_{01}. 
    \end{aligned} 
    \label{eq:app_transition_frequencies} 
\end{equation}
Although the eigenfrequencies in Table~\ref{tab:transmon_parameters} are expressed in units of $\omega_C$, the gate times and
filter bandwidths in the main text are reported using the directly measurable scale $|\alpha_2|$. The dimensionless quantities
displayed in the figures are therefore $\abs{\alpha_2}t_\uf$ and $\omega_{\mm{BW}}/\abs{\alpha_2}$, respectively.

\section{Grid search over the DRAG coefficient}
\label{app:drag_scan}

Leading-order DRAG is used in the main text as the practical low-complexity benchmark for compressed Magnus control.  For the
uncorrected in-phase envelope $f_x(t)$, the DRAG quadrature correction is
\begin{equation}
    g_{\mm{DRAG},y}(t;\beta) = -\frac{\beta}{\alpha_2} \dot{f}_x (t),
    \label{eq:app_drag_correction}
\end{equation}
where $\alpha_2=\omega_{12}-\omega_{01}$ is the anharmonicity associated with the first leakage level and $\beta$ is a
dimensionless coefficient.

The value of $\beta$ yielding the smallest average fidelity error can depend on the dimensionless gate time $\abs{\alpha_2}t_\uf$.
Equation~\eqref{eq:app_drag_correction} is a leading-order correction derived from a three-level Duffing approximation, whereas
the reported average fidelity error is obtained by propagating the complete four-level transmon model. Changing the gate duration
changes the relative contributions of leakage, phase errors, and higher-order dynamics that are not captured by the leading-order
DRAG expression.

We determine the DRAG benchmark in the main text using a discrete grid search. At each displayed gate time, we evaluate
\begin{equation}
    \beta \in D = \left\{0,0.1,0.2,0.3,0.4,0.5,0.6,0.7,0.8,0.9,1.0 \right\}.
    \label{eq:app_drag_beta_grid}
\end{equation}
For every pair $(\abs{\alpha_2} t_\uf,\beta)$, we construct the quadrature correction from Eq.~\eqref{eq:app_drag_correction},
numerically propagate the same four-level transmon Hamiltonian used in the main text, and compute the average fidelity error from
the projected final-time evolution.  No continuous optimization over $\beta$ is performed.

\begin{figure}[b]
    \centering
    \includegraphics[width=0.6\linewidth]{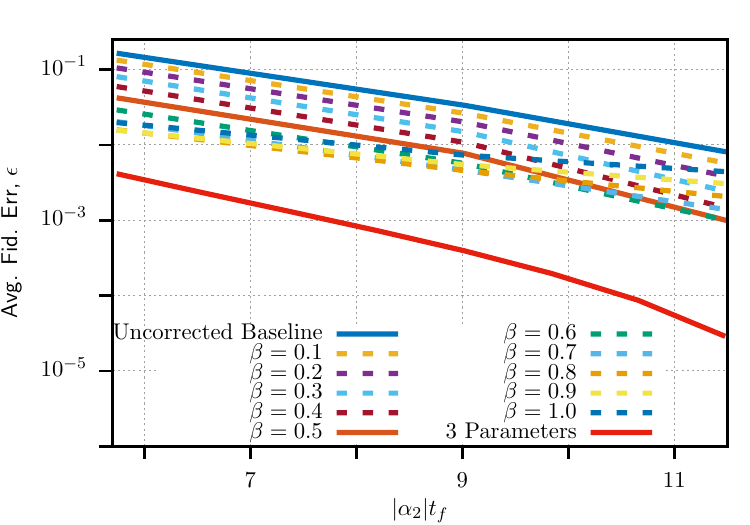}
    \caption{
        Average fidelity error $\epsilon$ versus dimensionless gate time $\abs{\alpha_2}t_\uf$ for the fixed DRAG coefficients
        $\beta=0.1,0.2,\ldots,1.0$, together with the uncorrected pulse and the three-parameter compressed Magnus correction.  For
        each curve, the four-level transmon Hamiltonian is propagated numerically and the final evolution is projected onto the
        computational subspace. The case $\beta=0$ coincides with the uncorrected pulse and is therefore not shown separately.
        The DRAG benchmark reported in Fig.~\ref{fig:performance}(c) is obtained by taking, at each gate time, the smallest
        average fidelity error among the sampled values of $\beta$.
    }
    \label{fig:app_drag_beta_scan}
\end{figure}

The DRAG error reported in Fig.~\ref{fig:performance}(c) is the pointwise minimum over this finite grid,
\begin{equation}
    \epsilon_{\mm{DRAG}}^{\mm{grid}} (\abs{\alpha_2} t_\uf) = \min_{ \beta\in D } \epsilon_{\mm{DRAG}}(\abs{\alpha_2}  t_\uf;\beta).
    \label{eq:app_drag_grid_envelope}
\end{equation}
Thus, $\epsilon_{\mm{DRAG}}^{\mm{grid}}(\abs{\alpha_2} t_\uf)$ is the lower envelope of the sampled fixed-$\beta$ curves.  It
should not be interpreted as the result of a continuous minimization over $\beta$, since a value between two grid points could in
principle yield a slightly smaller error.  Nevertheless, the grid-selected envelope provides a substantially stronger benchmark
than using a single fixed DRAG coefficient over the entire gate-time range.

Figure~\ref{fig:app_drag_beta_scan} shows that the preferred DRAG coefficient varies with $\abs{\alpha_2}t_\uf$:~No single fixed
value of $\beta$ yields the smallest error over the full displayed range. The three-parameter compressed Magnus correction
nevertheless produces a smaller average fidelity error than every sampled fixed-$\beta$ DRAG curve throughout this interval.
Consequently, it also outperforms the grid-selected DRAG envelope $\epsilon_{\mm{DRAG}}^{\mm{grid}}(\abs{\alpha_2} t_\uf)$ used in the main text.

This comparison is not intended to match the number of adjustable coefficients in the two constructions. Leading-order DRAG is
included because it provides the established experimental baseline for suppressing leakage with a simple waveform.  More elaborate
DRAG extensions and generic numerical optimal-control pulses introduce additional model or waveform complexity and therefore
constitute different benchmarks.

\section{Iterative fourth-order Magnus correction for an ill-conditioned correction Hamiltonian}
\label{app:fourth_order_iterative}

This section describes how we compute the fully parameterized fourth-order Magnus correction used as the reference pulse in
Fig.~\ref{fig:performance}. The procedure follows the singular or ill-conditioned construction of Ref.~\cite{Roque2021NPJ}. Rather
than solving for the fourth-order effective correction in one step, which would require solving a large system of coupled
fourth-order polynomial equations for all control parameters, we construct the correction iteratively. We first determine a
second-order effective correction $\hW_{\mm{eff},2}(t)$ and then obtain an additional fourth-order correction
$\hW_{\mm{eff},4}(t)$ around this fixed lower-order solution.

For the driven transmon, the physical lab-frame correction is applied through the charge-drive channel,
\begin{equation}
    \hW (t) = \left[ g_x(t) \cos\left(\omega_\ud t \right) + g_y (t) \sin\left(\omega_\ud t\right) \right]\hn,
    \label{eq:app_lab_control}
\end{equation}
where $g_x (t)$ and $g_y (t)$ are the correction envelopes. We keep the drive frequency constant throughout the pulse and write
$\omega_\ud = \omega_{01} + \Delta$. The constant detuning $\Delta$ therefore provides an additional control parameter in the
rotating frame without requiring a chirped carrier. The unknown control parameters are the Fourier coefficients of the physical
envelopes $g_x (t)$ and $g_y (t)$, together with the constant detuning $\Delta$.

The operator basis used in the cancellation equations is introduced after transforming to the frame rotating at the drive
frequency and truncating to the four lowest transmon eigenstates:~The two computational states and two leakage states; see
Ref.~\cite{Stanescu2026}. In this finite-dimensional description, the charge operator couples several pairs of transmon
eigenstates, with relative strengths fixed by the matrix elements $n_{ij}=\bra{i}\hn\ket{j}$. After transforming to the
interaction picture, we decompose the resulting Hamiltonian and Magnus generator in a convenient Hermitian operator basis
$\{\hA_\mu\}$ on the truncated Hilbert space. For example, this basis may include the generalized Pauli operators
\begin{equation}
    \begin{aligned}
        \hat{\sigma}_x^{(i,j)} &= \ketbra{i}{j} + \ketbra{j}{i},\\
        \hat{\sigma}_y^{(i,j)} &= -i\ketbra{i}{j} + i\ketbra{j}{i},\\
        \hat{\sigma}_z^{(i,j)} &= \ketbra{i}{i} - \ketbra{j}{j}.
    \end{aligned}
    \label{eq:app_generalized_paulis}
\end{equation}
These operators are not independently addressable lab-frame controls. They form a bookkeeping basis for identifying the
computational errors and leakage couplings generated collectively by the physical charge drive and its Magnus commutators.

The physical correction envelopes are expanded in the Fourier basis used in the main text,
\begin{equation}
    g_j (t) = \sum_{m=1}^{M_j} a_{j,m} \left[ 1 - \cos\left(2\pi m\frac{t}{t_\uf}\right) \right] +
    \sum_{l=1}^{L_j} b_{j,l} \sin\left(2\pi l\frac{t}{t_\uf}\right), \quad j=x,y.
    \label{eq:app_fourier_envelopes}
\end{equation}
The basis functions vanish at the beginning and end of the protocol. The Fourier coefficients determine the time dependence of the
two correction envelopes, not independent amplitudes for individual operators in $\{\hA_\mu\}$. Each envelope multiplies the
complete charge operator, which in the truncated transmon eigenbasis is given by Eq.~\eqref{eq:app_charge_operator_decomposition}.
Consequently, changing one Fourier coefficient generally modifies several transition channels simultaneously, with their relative
coupling strengths fixed by the charge-matrix elements $n_{ij}$. The operator basis $\{\hA_\mu\}$ is used only to decompose the
resulting interaction-picture Hamiltonian and Magnus generator into scalar cancellation equations.

We next specify the error components canceled by the fully parameterized correction. Let $\hP$ project onto the computational
subspace and let $\hQ = \mathbbm{1} - \hP$ project onto the two leakage levels. The cancellation target is fixed by these
projectors:~It consists of the traceless error acting within $\hP$ and all couplings between $\hP$ and $\hQ$. A component
proportional to $\hP$ generates only a common phase on the computational subspace, while a block acting entirely within $\hQ$ does
not affect a state initially prepared in $\hP$ without an accompanying coupling between the two subspaces. We therefore define the
fixed projection
\begin{equation} 
    \mathcal{P}_{\mm{rel}}[\hO] = \hP\hO\hP + \hP\hO\hQ + \hQ\hO\hP - \frac{ \Tr\left( \hP\hO\hP \right) }{ \Tr\hP } \hP. 
    \label{eq:app_relevant_projection} 
\end{equation} 
Here, the label ``rel'' refers to this fixed projection associated with the computational subspace; it does not denote a subset of
error channels selected during the calculation. We denote by $B_{\hP}$ a complete Hermitian operator basis for the image of
$\mathcal{P}_{\mm{rel}}$. Thus, the fully parameterized construction cancels every independent component of the projected
diagnostic generator in $B_{\hP}$, including all phase errors within the computational subspace and all leakage couplings from it.

For the present four-level model, the computational subspace is spanned by $\{\ket{0},\ket{1}\}$ and the leakage subspace by
$\{\ket{2},\ket{3}\}$. The traceless computational-subspace block is spanned by $\hat{\sigma}_x^{(0,1)}$,
$\hat{\sigma}_y^{(0,1)}$, and $\hat{\sigma}_z^{(0,1)}$, giving three independent coherent-error operators. The coupling between
the computational and leakage subspaces contains the four transitions $0\leftrightarrow2$, $0\leftrightarrow3$,
$1\leftrightarrow2$, and $1\leftrightarrow3$. Each transition contributes the two independent Hermitian operators
$\hat{\sigma}_x^{(i,j)}$ and $\hat{\sigma}_y^{(i,j)}$. The complete projected diagnostic generator is therefore spanned by $3 + 2
\times 4 =11$ independent error operators:~Three describing coherent errors within the computational subspace and eight describing
couplings between the computational and leakage subspaces. The common phase on the computational subspace and operators acting
entirely within the leakage subspace are excluded by the projection in Eq.~\eqref{eq:app_relevant_projection}.

The $11$ error operators define the cancellation conditions and should not be confused with the number of independently adjustable
waveform parameters required to satisfy them. In the present constrained-control problem, some error operators are generated only
through higher-order Magnus commutators, so the cancellation equations depend nonlinearly on the waveform coefficients. Satisfying
these equations can therefore require more independently adjustable control parameters than there are projected error components.

We now describe the iterative construction. Let $\hV_\uI(t)$ denote the interaction-picture error Hamiltonian obtained from the
uncorrected drive. Because the available charge drive does not directly span all operator directions in $B_{\hP}$, the
regular linear Magnus construction is ill conditioned. In particular, some couplings involving the second leakage level $\ket{3}$
appear only through commutators of the available drive terms. We therefore first solve a second-order effective problem by
defining
\begin{equation}
    \hH_{\uI,\mm{eff}}^{(2)}(t) = \varepsilon \hV_\uI (t) + \hW_{\mm{eff},2,\uI}(t; \cv_2),
    \label{eq:app_H_eff2}
\end{equation}
computing its Magnus generator through second order, and forming the Hermitian diagnostic generator
\begin{equation}
    \hE_{\mm{eff}}^{(2)} (t_\uf;\cv_2) = -i \hOmega_{\mm{eff}}^{(2)} (t_\uf;\cv_2).
    \label{eq:app_E_eff2}
\end{equation}
Applying the fixed projection in Eq.~\eqref{eq:app_relevant_projection}, we obtain the complete second-order projected diagnostic
generator
\begin{equation} 
    \begin{aligned} 
        \hE_{\mm{rel},\mm{eff}}^{(2)} (t_\uf;\cv_2) &= \mathcal{P}_{\mm{rel}} \left[ \hE_{\mm{eff}}^{(2)} (t_\uf;\cv_2) \right] \\ 
        &= \sum_{\hA_\nu\in B_{\hP}} E_\nu^{(2)} (t_\uf;\cv_2) \hA_\nu. 
    \end{aligned}
    \label{eq:app_second_residual} 
\end{equation}
The second-order effective correction is determined by canceling every independent component of this projected generator,
\begin{equation} 
    E_\nu^{(2)}(t_\uf;\cv_2) = 0, \quad \hA_\nu \in B_{\hP}. 
    \label{eq:app_second_order_equations}
\end{equation} 
Because operator directions absent from the direct charge drive are generated through second-order commutators, the functions
$E_\nu^{(2)}$ are generally polynomial in the control parameters $\cv_2$. We denote a solution of
Eq.~\eqref{eq:app_second_order_equations} by $\cv_2^\star$ and the corresponding correction by $\hW_{\mm{eff},2}(t) =
\hW_{\mm{eff},2}(t;\cv_2^\star)$.

The fourth-order correction is obtained recursively. Directly solving for all control parameters of a fourth-order effective
correction would require a large coupled system of fourth-order polynomial equations. Instead, following the arbitrary-order
construction of Ref.~\cite{Roque2021NPJ}, we use the fixed second-order solution to define
\begin{equation}
    \hH_{\uI,\mm{eff}}^{(4)}(t) = \varepsilon \hV_\uI (t) + \hW_{\mm{eff},2,\uI} (t;\cv_2^\star) + \hW_{\mm{eff},4,\uI} (t;\cv_4).
    \label{eq:app_H_eff4}
\end{equation}
Here, $\hW_{\mm{eff},2,\uI} (t;\cv_2^\star)$ is the fixed second-order correction, whereas $\hW_{\mm{eff},4,\uI}(t;\cv_4)$ is the
additional correction to be determined at fourth order. The new correction is implemented using the same physical control
resources:~The envelopes $g_x(t)$ and $g_y(t)$ and the constant detuning $\Delta$.

We compute the Magnus expansion generated by Eq.~\eqref{eq:app_H_eff4} through fourth order, 
\begin{equation}
    \hOmega_{\mm{eff}}^{(4)}(t_\uf;\cv_4) = \sum_{k=1}^{4} \hOmega_{\mm{eff},k}^{(4)} (t_\uf;\cv_4).
    \label{eq:app_Omega_eff4}
\end{equation}
Following the approximation used in the arbitrary-order singular construction of Ref.~\cite{Roque2021NPJ}, we keep from this
fourth-order Magnus generator only terms containing at most two powers of the new correction $\hW_{\mm{eff},4,\uI}(t;\cv_4)$.
Equivalently, we discard terms containing three or more powers of $\hW_{\mm{eff},4,\uI}(t;\cv_4)$. The resulting cancellation
equations are therefore at most quadratic in the unknown parameters $\cv_4$, rather than fourth-order polynomial equations.

This approximation does not mean that only second-order errors are canceled. The background Hamiltonian $\varepsilon\hV_\uI(t) +
\hW_{\mm{eff},2,\uI}(t;\cv_2^\star)$ remains inside the Magnus expansion through fourth order, and $\hW_{\mm{eff},4,\uI}(t;\cv_4)$
is chosen to cancel, through order $\varepsilon^4$, every independent component of the projected diagnostic generator in
$B_{\hP}$. Terms quadratic in $\hW_{\mm{eff},4,\uI}(t;\cv_4)$ are essential in the ill-conditioned case because their commutators
generate operator directions that are absent from the direct charge-drive Hamiltonian.

We define the approximate fourth-order Hermitian diagnostic generator as 
\begin{equation} 
    \hE_{\mm{eff}}^{(4)} (t_\uf;\cv_4;\cv_2^\star) = -i \left. \hOmega_{\mm{eff}}^{(4)} (t_\uf;\cv_4;\cv_2^\star) \right|_{ \leq
        2\,\mm{powers\,of}\; \hW_{\mm{eff},4} }, 
    \label{eq:app_E_eff4} 
\end{equation}
where the vertical bar denotes the truncation just described. Applying the same fixed projection as at second order gives the
complete fourth-order projected diagnostic generator
\begin{equation} 
    \begin{aligned} 
        \hE_{\mm{rel},\mm{eff}}^{(4)} (t_\uf;\cv_4;\cv_2^\star) &= \mathcal{P}_{\mm{rel}} \left[ \hE_{\mm{eff}}^{(4)} (t_\uf;\cv_4;\cv_2^\star) \right] \\ 
        &= \sum_{\hA_\nu\in B_{\hP}} E_\nu^{(4)} (t_\uf;\cv_4;\cv_2^\star) \hA_\nu . 
    \end{aligned} 
    \label{eq:app_fourth_residual} 
\end{equation}
The fourth-order correction is determined by canceling every independent component of this projected generator, 
\begin{equation} 
    E_\nu^{(4)} (t_\uf;\cv_4;\cv_2^\star) = 0, \quad \hA_\nu\in B_{\hP}. 
    \label{eq:app_fourth_equations}
\end{equation} 
The system in Eq.~\eqref{eq:app_fourth_equations} is nonlinear but at most quadratic in $\cv_4$. This is the practical advantage
of the iterative construction:~The coefficients $\cv_2^\star$ have already been chosen so that
\begin{equation} 
    \hE_{\mm{rel},\mm{eff}}^{(2)} (t_\uf;\cv_2^\star) = \mathbf{0}. 
    \label{eq:app_second_generator_zero}
\end{equation}
Thus, every traceless error within the computational subspace and every coupling between the computational and leakage subspaces is
canceled through second order. The solution $\cv_4^\star$ then cancels every component of the same projected diagnostic generator
through fourth order, without requiring a one-shot solution of the full fourth-order polynomial problem.

After solving Eq.~\eqref{eq:app_fourth_equations}, the bookkeeping parameter $\varepsilon$ is set to one. In the decomposition of
Eq.~\eqref{eq:Hsplit}, the fourth-order correction is 
\begin{equation} 
    \hW^{(4)}(t) = \hW_{\mm{eff},2}(t;\cv_2^\star) + \hW_{\mm{eff},4}(t;\cv_4^\star). 
    \label{eq:app_W_corr4}
\end{equation} 
For the transmon drive, this correction is implemented through the two charge-drive quadratures and a constant detuning,
\begin{equation}
    \begin{aligned}
        \hW^{(4)} (t) &= \left[ g_x^{(4)} (t) \cos\left(\omega_\ud^{(4)} t \right) + g_y^{(4)} (t)
        \sin\left(\omega_\ud^{(4)} t\right) \right]\hn,\\
        \omega_\ud^{(4)} & = \omega_{01} + \Delta^{(4)}.
    \end{aligned}
    \label{eq:app_W_corr_quadratures}
\end{equation}
The drive frequency is held constant throughout the pulse. The correction is therefore specified by the two correction envelopes
$g_x^{(4)}(t)$ and $g_y^{(4)}(t)$ and the constant detuning $\Delta^{(4)}$. The correction envelopes are added to the
corresponding uncorrected drive quadratures,
\begin{equation}
    \begin{aligned}
        f_{x,\mm{mod}}^{(4)} (t) &= f_x (t) + g_x^{(4)}(t),\\
        f_{y,\mm{mod}}^{(4)} (t) &= f_y (t) + g_y^{(4)}(t).
    \end{aligned}
    \label{eq:app_modified_quadratures}
\end{equation}
The triplet $\{f_{x,\mm{mod}}^{(4)} (t), f_{y,\mm{mod}}^{(4)} (t), \Delta^{(4)}\}$ defines the fully parameterized fourth-order
modified pulse used as the reference in Figs.~\ref{fig:performance} and~\ref{fig:pulses}.

For the transmon calculation in the main text, the fully parameterized reference correction contains $17$ independently adjustable
control parameters:~One constant detuning and $16$ Fourier amplitudes, with four sine and four cosine coefficients for each
microwave quadrature. These $17$ control parameters should not be confused with the $11$ independent components of the projected
diagnostic generator. The latter define the cancellation conditions, whereas the former provide the waveform degrees of freedom
used to solve those coupled nonlinear conditions.

The need for additional control freedom arises because some projected error operators are generated only through higher-order
Magnus commutators. Canceling an indirectly generated error channel can therefore require independent adjustment of several
intermediate contributions entering the corresponding commutator pathways. For example, controlling an effective $0\to 3$
contribution requires sufficient independent control of contributions involving the $0\to 2$ and $2\to 3$ transitions. The
physical envelopes do not address these transitions independently; rather, the Fourier coefficients modify the complete
charge-drive waveform and thereby provide the control freedom needed to satisfy all $11$ nonlinear cancellation conditions through
fourth order. The same Fourier counting gives the fourth harmonic discussed in the finite-bandwidth analysis of
Sec.~\ref{app:filtering}.

\section{Derivation and properties of the compressed Magnus functional}
\label{app:compressed_functional}

This section derives the spectral functional used in the compressed Magnus construction and clarifies its dependence on the
retained Fourier coefficients. At perturbative order $n$, the interaction-picture evolution is approximated by

\begin{equation}
    \hU_\uI^{(n)} (t_\uf;\cv^{(n)}) = \exp\left[\hOmega^{(n)} (t_\uf; \cv^{(n)}) \right],
    \label{eq:app_UI_Magnus}
\end{equation}
where $\cv^{(n)}$ denotes the retained Fourier coefficients at order $n$.  The truncated Magnus generator
$\hOmega^{(n)} (t_\uf; \cv^{(n)})$ is anti-Hermitian, and we define the Hermitian diagnostic error generator
\begin{equation}
    \hE^{(n)} (t_\uf;\cv^{(n)} ) = -i \hOmega^{(n)} (t_\uf;\cv^{(n)}).
    \label{eq:app_Edef}
\end{equation}
The operator $\hE^{(n)}$ is not a physical Hamiltonian. Its eigenvalues are dimensionless eigenphases of the truncated residual
interaction-picture evolution, and perturbation theory requires the projected residual eigenphases to remain small.

We isolate the error components associated with the target subspace using the fixed superoperator $\mathcal{P}_{\mm{rel}}$ defined
in Eq.~\eqref{eq:app_relevant_projection}. This projection retains the traceless block acting within $\hP$ and the blocks coupling
$\hP$ and $\hQ$, while removing the common phase on $\hP$ and discarding dynamics entirely internal to $\hQ$. Applying it to the
truncated diagnostic generator gives
\begin{equation} 
    \hE_{\mm{rel}}^{(n)} (t_\uf;\cv^{(n)}) = \mathcal{P}_{\mm{rel}} \left[ \hE^{(n)} (t_\uf;\cv^{(n)}) \right]. 
    \label{eq:app_relevant_generator} 
\end{equation} 

The compressed Magnus cost is the scalar function
\begin{equation}
    \Phi^{(n)} (\cv^{(n)}) = \Tr\left( \exp\left[ \hE_{\mm{rel}}^{(n)} (\cv^{(n)}) \right] + \exp\left[
    -\hE_{\mm{rel}}^{(n)} (\cv^{(n)}) \right] - 2\mathbbm{1}_{\mm{supp}} \right),
    \label{eq:app_Phi}
\end{equation}
where $\mathbbm{1}_{\mm{supp}}$ denotes the identity on the support of $\hE_{\mm{rel}}^{(n)}$. Because
$\hE_{\mm{rel}}^{(n)} (\cv^{(n)})$ is Hermitian, it has a spectral decomposition
\begin{equation}
    \hE_{\mm{rel}}^{(n)} (\cv^{(n)} ) = \sum_j \lambda_j (\cv^{(n)}) \ketbra{\lambda_j
    (\cv^{(n)})}{\lambda_j (\cv^{(n)})}.
    \label{eq:app_spectral_decomp}
\end{equation}
Here, $\lambda_j (\cv^{(n)})$ are the real eigenvalues of the projected diagnostic generator, and $\ket{\lambda_j (\cv^{(n)})}$
are the corresponding orthonormal eigenstates on the support of $\hE_{\mm{rel}}^{(n)}$.  These eigenvalues are the residual
eigenphases penalized by the compressed Magnus functional.

Using the spectral decomposition, the functional becomes
\begin{equation}
    \begin{aligned}
        \Phi^{(n)} (\cv^{(n)}) &= \Tr\left( \exp\left[ \hE_{\mm{rel}}^{(n)} (\cv^{(n)}) \right] + \exp\left[
        -\hE_{\mm{rel}}^{(n)} (\cv^{(n)}) \right] - 2\mathbbm{1}_{\mm{supp}} \right) \\
        &= \sum_j \left(\exp\left[\lambda_j (\cv^{(n)}) \right] + \exp\left[ -\lambda_j (\cv^{(n)}) \right] - 2 \right) \\
        &= 2 \sum_j \left( \cosh\left[\lambda_j (\cv^{(n)})\right] - 1 \right).
    \end{aligned}
    \label{eq:app_PhiEig}
\end{equation}
This expression shows that $\Phi^{(n)} (\cv^{(n)}) \geq 0$ and that it is minimized when all relevant residual
eigenphases vanish.  It also makes explicit why the functional is tied to the diagnostic generator:~The optimization penalizes the
eigenphases of the truncated interaction-picture error, not a final-time fidelity.

Near the perturbative origin, the functional reduces to the Frobenius norm squared of the relevant diagnostic generator.
Expanding the exponentials gives
\begin{equation}
    \exp\left[ \hE_{\mm{rel}}^{(n)} (\cv^{(n)})   \right] + \exp\left[ -\hE_{\mm{rel}}^{(n)} (\cv^{(n)}) \right] - 2\mathbbm{1}_{\mm{supp}}
    = \left[\hE_{\mm{rel}}^{(n)} (\cv^{(n)}) \right]^2 + \frac{1}{12} \left[\hE_{\mm{rel}}^{(n)} (\cv^{(n)}) \right]^4
    + \BigO\left[ \left( \hE_{\mm{rel}}^{(n)} (\cv^{(n)}) \right)^6 \right].
    \label{eq:app_matrix_expansion}
\end{equation}
Taking the trace,
\begin{equation}
    \begin{aligned}
        \Phi^{(n)}(\cv^{(n)}) &= \Tr\left[ \left( \hE_{\mm{rel}}^{(n)}(\cv^{(n)}) \right)^2 \right] + \frac{1}{12} \Tr\left[ \left( \hE_{\mm{rel}}^{(n)}(\cv^{(n)}) \right)^4 \right] +
        \BigO\left[ \norm{ \hE_{\mm{rel}}^{(n)}(\cv^{(n)}) }_\uF^6 \right] \\
        &= \norm{\hE_{\mm{rel}}^{(n)}(\cv^{(n)}) }_\uF^2 + \BigO\left[ \norm{ \hE_{\mm{rel}}^{(n)}(\cv^{(n)}) }_\uF^4 \right],
    \end{aligned}
    \label{eq:app_Frobenius}
\end{equation}
where $\lVert \hO\rVert_\uF= \Tr[\hO \hO^\dagger]^{1/2}$ is the Frobenius norm. Thus, locally, minimizing $\Phi^{(n)}(\cv^{(n)}) $
is equivalent to minimizing the Frobenius norm of the projected truncated diagnostic generator $\hE_{\mm{rel}}^{(n)}$. The
higher-order even powers penalize large eigenvalues of either sign, i.e., large $|\lambda_j|$, and therefore disfavor
nonperturbatively large residual eigenphases.

Finally, we state the convexity property in the notation used for the control problem. The map
\begin{equation}
    \hO \mapsto \Tr\left[ \exp\left( \hO \right) + \exp\left( -\hO \right) - 2\mathbbm{1} \right]
\end{equation}
is a convex spectral function of the Hermitian operator $\hO$. Therefore, $\Phi^{(n)}$ is convex with respect to the Hermitian
diagnostic generator $\hE_{\mm{rel}}^{(n)}$.

This convexity is a statement about the generator, not automatically about the Fourier coefficients. The actual control variables
are the retained coefficients $\cv^{(n)}$, and the cost is the composition $\Phi^{(n)}(\cv^{(n)}) =
\Phi^{(n)}[\hE_{\mm{rel}}^{(n)}(\cv^{(n)})]$. Convexity is preserved under affine changes of variables, but not under a
general nonlinear map from coefficients to generator. Thus, if
\begin{equation}
    \hE_{\mm{rel}}^{(n)}(\cv^{(n)}) = \hE_0 + \sum_\mu c_\mu^{(n)} \hE_\mu .
    \label{eq:app_linear_dependence}
\end{equation}
the coefficient-space problem inherits convexity from the generator objective. In singular or ill-conditioned cases, missing
operator directions are generated through higher-order Magnus commutators, and the dependence is generally polynomial,
\begin{equation}
    \hE_{\mm{rel}}^{(n)}(\cv^{(n)}) = \hE^{(0)} + \sum_\mu c_\mu^{(n)} \hE_\mu^{(1)} + \sum_{\mu\nu} c_\mu^{(n)}c_\nu^{(n)} \hE_{\mu\nu}^{(2)} + \cdots .
    \label{eq:app_polynomial_dependence}
\end{equation}
Global convexity in the Fourier coefficients is then not guaranteed. The essential point is that the objective remains a
physically meaningful spectral penalty that is convex with respect to the analytically computed diagnostic generator, even when
its pullback to the coefficient space is nonconvex. At the generator level, its unique minimum corresponds to vanishing residual
eigenphases.

\section{Relation to the average fidelity error}
\label{sec:fidelity_relation}

We now relate the compressed Magnus functional to the average fidelity error used in the numerical simulations.  Let
$\hU_{\mm{ideal}}$ denote the target unitary on the $d$-dimensional subspace of interest, where $d=\Tr\hP$.  We regard
$\hU_{\mm{ideal}}$ as an operator supported on this subspace, so that $\hU_{\mm{ideal}}=\hP\hU_{\mm{ideal}}\hP$.  The projected
error operator is
\begin{equation}
    \hO (t_\uf) = \hU_{\mm{ideal}}^\dagger \hP \hU(t_\uf) \hP.
    \label{eq:sup_error_operator}
\end{equation}
The corresponding average fidelity error is
\begin{equation}
    \epsilon = 1 - \frac{1}{d(d+1)} \left[ \Tr\left( \hO\hO^\dagger \right) + \abs{ \Tr\left( \hO \right) }^2 \right].
    \label{eq:sup_average_infidelity}
\end{equation}
Because the projected operator $\hO$ need not be unitary when leakage is present, Eq.~\eqref{eq:sup_average_infidelity} accounts
for both coherent errors within the subspace of interest and population transferred to its complement.

The total evolution can be decomposed into the ideal and interaction-picture evolutions,
\begin{equation}
    \hU (t_\uf) = \hU_0 (t_\uf) \hU_\uI (t_\uf).
    \label{eq:sup_total_evolution}
\end{equation}
By construction, the restriction of the ideal evolution to the subspace of interest equals the target operation,
\begin{equation}
    \hP\hU_0 (t_\uf) \hP = \hU_{\mm{ideal}}.
    \label{eq:sup_ideal_evolution}
\end{equation}
In addition, the ideal evolution does not couple the subspace of interest and its complement:
\begin{equation}
    \begin{aligned}
        \hQ \hU_0 (t_\uf) \hP &= \mathbf{0}, \\ 
        \hP \hU_0 (t_\uf)\hQ &= \mathbf{0}.
    \end{aligned}
    \label{eq:sup_ideal_decoupling}
\end{equation}
This is what is meant by saying that $\hU_0(t_\uf)$ preserves the subspace of interest.  It does not imply that $\hU_0(t_\uf)$ has
support only on that subspace; $\hU_0(t_\uf)$ may also generate nontrivial dynamics within the complementary subspace, but it is
block diagonal with respect to the decomposition $\mathbbm{1}=\hP+\hQ$.

Using Eqs.~\eqref{eq:sup_total_evolution} and \eqref{eq:sup_ideal_evolution}, and inserting $\mathbbm{1}=\hP+\hQ$ between
$\hU_0(t_\uf)$ and $\hU_\uI(t_\uf)$, gives
\begin{equation}
    \begin{aligned}
        \hO(t_\uf) &= \hU_{\mm{ideal}}^\dagger \hP\hU_0(t_\uf) \left( \hP+\hQ \right) \hU_\uI(t_\uf)\hP \\
        &= \hU_{\mm{ideal}}^\dagger \hP\hU_0(t_\uf)\hP \hU_\uI(t_\uf)\hP + \hU_{\mm{ideal}}^\dagger \hP\hU_0(t_\uf)\hQ \hU_\uI(t_\uf)\hP \\
        &= \hU_{\mm{ideal}}^\dagger \hU_{\mm{ideal}} \hP\hU_\uI(t_\uf)\hP \\
        &= \hP\hU_\uI(t_\uf)\hP.  
    \end{aligned}
    \label{eq:sup_projected_interaction_evolution}
\end{equation}
In the last two lines, the term containing $\hP\hU_0(t_\uf)\hQ$ vanishes by Eq.~\eqref{eq:sup_ideal_decoupling}, and we used the
fact that $\hU_{\mm{ideal}}$ is unitary on the subspace of interest, so that $\hU_{\mm{ideal}}^\dagger\hU_{\mm{ideal}}=\hP$.

After optimizing the retained coefficients, the truncated interaction-picture evolution is
\begin{equation}
    \hU_\uI^{(n)} \left( t_\uf; \cv_\star^{(n)} \right) = \exp\left[ i \hE^{(n)} \left( t_\uf; \cv_\star^{(n)} \right) \right],
    \label{eq:sup_UI_error_generator}
\end{equation}
where we used $\hE^{(n)}=-i\hOmega^{(n)}$, and hence $\hOmega^{(n)}=i\hE^{(n)}$.  Equation~\eqref{eq:sup_UI_error_generator} is
understood up to corrections beyond the retained Magnus order.

For compactness, we define
\begin{equation}
    \begin{aligned}
        \hE_\star &= \hE^{(n)} \left( t_\uf; \cv_\star^{(n)} \right), \\ 
        \hE_{PP} &= \hP\hE_\star\hP, \\
        \hE_{QP} &= \hQ\hE_\star\hP, \\
        \hE_{PQ} &= \hP\hE_\star\hQ = \hE_{QP}^\dagger .
    \end{aligned}
    \label{eq:sup_generator_blocks}
\end{equation}
The operator $\hE_\star$ is the residual diagnostic generator after optimizing the retained control coefficients. It generates the
deviation of the truncated interaction-picture evolution from the identity. It is decomposed into an operator $\hE_{PP}$
describing coherent errors within the subspace of interest, and an operator $\hE_{QP}$ describing the coupling from that subspace
to its complement. The perturbative construction is controlled when the relevant eigenvalues of $\hE_\star$ are small.

Expanding the interaction-picture evolution to second order in the residual generator gives
\begin{equation}
    \hU_\uI^{(n)} \left( t_\uf; \cv_\star^{(n)} \right) = \mathbbm{1} + i\hE_\star - \frac{1}{2}\hE_\star^2 + \BigO\left( \norm{\hE_\star}^3 \right).
    \label{eq:sup_UI_expansion}
\end{equation}
Projecting this expression onto the subspace of interest yields
\begin{equation}
    \hO = \hP + i\hE_{PP} - \frac{1}{2} \hP\hE_\star^2\hP + \BigO\left( \norm{\hE_\star}^3 \right).
    \label{eq:sup_O_preliminary_expansion}
\end{equation}
The quadratic term can be decomposed by inserting $\mathbbm{1}=\hP+\hQ$ between the two factors of $\hE_\star$:
\begin{equation}
    \begin{aligned}
        \hP\hE_\star^2\hP &= \hP\hE_\star \left( \hP+\hQ \right) \hE_\star\hP \\
        &= \hP\hE_\star\hP \hE_\star\hP + \hP\hE_\star\hQ \hE_\star\hP \\
        &= \hE_{PP}^2 + \hE_{PQ}\hE_{QP} \\ 
        &= \hE_{PP}^2 + \hE_{QP}^\dagger\hE_{QP}.  
    \end{aligned}
    \label{eq:sup_quadratic_block_decomposition}
\end{equation}
Consequently,
\begin{equation}
    \hO = \hP + i\hE_{PP} - \frac{1}{2} \left( \hE_{PP}^2 + \hE_{QP}^\dagger\hE_{QP} \right) + \BigO\left( \norm{\hE_\star}^3 \right).
    \label{eq:sup_error_operator_expansion}
\end{equation}
Taking the Hermitian conjugate gives
\begin{equation}
    \hO^\dagger = \hP - i\hE_{PP} - \frac{1}{2} \left( \hE_{PP}^2 + \hE_{QP}^\dagger\hE_{QP} \right) + \BigO\left( \norm{\hE_\star}^3 \right).
    \label{eq:sup_error_operator_dagger}
\end{equation}
Multiplying Eqs.~\eqref{eq:sup_error_operator_expansion} and \eqref{eq:sup_error_operator_dagger}, and retaining terms through
second order, gives
\begin{equation}
    \begin{aligned}
        \hO\hO^\dagger &= \hP - \left( \hE_{PP}^2 + \hE_{QP}^\dagger\hE_{QP} \right) + \hE_{PP}^2 + \BigO\left( \norm{\hE_\star}^3 \right) \\
        &= \hP - \hE_{QP}^\dagger\hE_{QP} + \BigO\left( \norm{\hE_\star}^3 \right).  \end{aligned}
    \label{eq:sup_OO_dagger}
\end{equation}
Thus,
\begin{equation}
    \Tr\left( \hO\hO^\dagger \right) = d - \Tr\left( \hE_{QP}^\dagger\hE_{QP} \right) + \BigO\left( \norm{\hE_\star}^3 \right).
    \label{eq:sup_survival_trace}
\end{equation}
The trace of Eq.~\eqref{eq:sup_error_operator_expansion} is
\begin{equation}
    \Tr\left( \hO \right) = d + i\Tr\left( \hE_{PP} \right) - \frac{1}{2} \Tr\left( \hE_{PP}^2 + \hE_{QP}^\dagger\hE_{QP} \right) + \BigO\left( \norm{\hE_\star}^3 \right).
    \label{eq:sup_trace_O}
\end{equation}
Since $\hE_{PP}$ is Hermitian, $\Tr(\hE_{PP})$ and $\Tr(\hE_{PP}^2)$ are real. Therefore,
\begin{equation}
    \abs{ \Tr\left( \hO \right) }^2 = d^2 - d\, \Tr\left( \hE_{PP}^2 + \hE_{QP}^\dagger\hE_{QP} \right) + \left[ \Tr\left( \hE_{PP} \right) \right]^2 + \BigO\left( \norm{\hE_\star}^3 \right).
    \label{eq:sup_trace_overlap}
\end{equation}
Substituting Eqs.~\eqref{eq:sup_survival_trace} and \eqref{eq:sup_trace_overlap} into Eq.~\eqref{eq:sup_average_infidelity} gives
\begin{equation}
    \epsilon = \frac{ d\, \Tr\left( \hE_{PP}^2 \right) - \left[ \Tr\left( \hE_{PP} \right) \right]^2 }{ d(d+1) }
    + \frac{1}{d} \Tr\left( \hE_{QP}^\dagger\hE_{QP} \right) + \BigO\left( \norm{\hE_\star}^3 \right).  
    \label{eq:sup_infidelity_generator}
\end{equation}
The first term is unchanged by adding a component proportional to $\hP$ to $\hE_{PP}$.  Such a component generates only a common
phase on the subspace of interest.  We therefore define the traceless computational-subspace block
\begin{equation}
    \hE_{PP,0} = \hE_{PP} - \frac{ \Tr\left( \hE_{PP} \right) }{ d } \hP.
    \label{eq:sup_traceless_PP}
\end{equation}
Its squared Frobenius norm is
\begin{equation}
    \norm{\hE_{PP,0}}_F^2 = \Tr\left( \hE_{PP}^2 \right) - \frac{ \left[ \Tr\left( \hE_{PP} \right) \right]^2 }{ d }.
    \label{eq:sup_traceless_PP_norm}
\end{equation}
Equation~\eqref{eq:sup_infidelity_generator} can therefore be written as
\begin{equation}
    \epsilon = \frac{1}{d+1} \norm{\hE_{PP,0}}_F^2 + \frac{1}{d} \norm{\hE_{QP}}_F^2 + \BigO\left( \norm{\hE_\star}^3 \right).  
    \label{eq:sup_infidelity_blocks}
\end{equation}
Thus, coherent errors within the subspace of interest and couplings from that subspace to its complement both contribute
positively to the leading average fidelity error. At this stage the remainder is written conservatively as
$\BigO(\norm{\hE_\star}^3)$. Below, the invariance of the fidelity under $\hE_\star\rightarrow-\hE_\star$ sharpens this remainder
to quartic order.

At the optimized coefficients, the relevant diagnostic generator defined in the main text is
\begin{equation}
    \hE_{\mm{rel},\star}^{(n)} = \hE_{PP,0} + \hE_{PQ} + \hE_{QP}.
    \label{eq:sup_relevant_generator}
\end{equation}
The three blocks in Eq.~\eqref{eq:sup_relevant_generator} are mutually orthogonal under the Hilbert--Schmidt inner product.  Since
$\hE_{PQ}=\hE_{QP}^\dagger$, its squared Frobenius norm is
\begin{equation}
    \begin{aligned}
        \norm{ \hE_{\mm{rel},\star}^{(n)} }_F^2 &= \norm{ \hE_{PP,0} }_F^2 + \norm{ \hE_{PQ} }_F^2 + \norm{ \hE_{QP} }_F^2 \\
        &= \norm{ \hE_{PP,0} }_F^2 + 2 \norm{ \hE_{QP} }_F^2.
    \end{aligned}
    \label{eq:sup_relevant_frobenius}
\end{equation}
The compressed Magnus cost satisfies
\begin{equation}
    \Phi^{(n)} \left( \cv_\star^{(n)} \right) = \norm{ \hE_{\mm{rel},\star}^{(n)} }_F^2 + \BigO\left( \norm{ \hE_{\mm{rel},\star}^{(n)} }_F^4 \right).
    \label{eq:sup_phi_frobenius}
\end{equation}
Equations~\eqref{eq:sup_infidelity_blocks} and \eqref{eq:sup_relevant_frobenius} provide the desired connection between generator
minimization and the average fidelity error.

To make this connection quantitative, we separate the leading quadratic contributions from the complete average fidelity error and
compressed Magnus functional:
\begin{equation}
    \begin{aligned}
        \epsilon &= \epsilon^{(2)} + \BigO\left( \norm{\hE_\star}^4 \right) + \delta\epsilon_{>n}, \\
        \Phi^{(n)} \left( \cv_\star^{(n)} \right) &= \Phi^{(n,2)} + \BigO\left( \norm{ \hE_{\mm{rel},\star}^{(n)} }_F^4 \right).
    \end{aligned}
    \label{eq:sup_full_and_quadratic_quantities}
\end{equation}
The quantity $\epsilon^{(2)}$ denotes the contribution to the average fidelity error that is quadratic in this residual generator,
while $\Phi^{(n,2)}$ denotes the quadratic contribution to the compressed Magnus functional constructed from the order-$n$
truncated generator. Thus, the first superscript in $\Phi^{(n,2)}$ specifies the retained Magnus order, whereas the second
specifies the power of the residual generator in the local expansion. The separate term $\delta\epsilon_{>n}$ in
Eq.~\eqref{eq:sup_full_and_quadratic_quantities} denotes corrections arising from terms beyond the order-$n$ truncated Magnus
expansion.

The absence of odd powers of $\hE_\star$ in the fidelity expansion follows from the invariance of
Eq.~\eqref{eq:sup_average_infidelity} under $\hE_\star\rightarrow-\hE_\star$. Indeed,
\begin{equation}
    \hP\exp\left( -i\hE_\star \right)\hP = \left[ \hP\exp\left( i\hE_\star \right)\hP \right]^\dagger ,
    \label{eq:sup_generator_sign_reversal}
\end{equation}
and both $\Tr(\hO\hO^\dagger)$ and $\abs{\Tr(\hO)}^2$ are unchanged under $\hO\rightarrow\hO^\dagger$.  The first correction to
$\epsilon^{(2)}$ arising from higher powers of the residual generator is therefore quartic.  

From Eq.~\eqref{eq:sup_infidelity_blocks}, the leading quadratic fidelity error is
\begin{equation}
    \epsilon^{(2)} = \frac{1}{d+1} \norm{ \hE_{PP,0} }_F^2 + \frac{1}{d} \norm{ \hE_{QP} }_F^2.
    \label{eq:sup_infidelity_quadratic}
\end{equation}
The first term measures coherent error within the subspace of interest after removing the irrelevant common phase, while the
second measures coupling from that subspace to its complement.

Similarly, Eqs.~\eqref{eq:sup_relevant_frobenius} and \eqref{eq:sup_phi_frobenius} give
\begin{equation}
    \begin{aligned}
        \Phi^{(n,2)} &= \norm{ \hE_{\mm{rel},\star}^{(n)} }_F^2 \\ &= \norm{ \hE_{PP,0} }_F^2 + 2 \norm{ \hE_{QP} }_F^2.
    \end{aligned}
    \label{eq:sup_functional_quadratic}
\end{equation}
Thus $\epsilon^{(2)}$ and $\Phi^{(n,2)}$ are positive quadratic forms of the same two physically relevant contributions. Their
relative weights differ, but each quadratic form bounds the other.

The lower bound follows directly from
\begin{equation}
    \begin{aligned}
        \epsilon^{(2)} - \frac{1}{2d} \Phi^{(n,2)} &= \left( \frac{1}{d+1} - \frac{1}{2d} \right) \norm{ \hE_{PP,0} }_F^2 \\ 
        &= \frac{d-1}{2d(d+1)} \norm{ \hE_{PP,0} }_F^2 \geq 0.
    \end{aligned}
    \label{eq:sup_infidelity_lower_bound_derivation}
\end{equation}
Similarly, the upper bound follows from
\begin{equation}
    \begin{aligned}
        \frac{1}{d+1} \Phi^{(n,2)} - \epsilon^{(2)} &= \left( \frac{2}{d+1} - \frac{1}{d} \right) \norm{ \hE_{QP} }_F^2 \\
        &= \frac{d-1}{d(d+1)} \norm{ \hE_{QP} }_F^2 \geq 0.
    \end{aligned}
    \label{eq:sup_infidelity_upper_bound_derivation}
\end{equation}
Consequently,
\begin{equation}
    \frac{1}{2d} \Phi^{(n,2)} \leq \epsilon^{(2)} \leq \frac{1}{d+1} \Phi^{(n,2)}.
    \label{eq:sup_infidelity_bounds_general}
\end{equation}
For the qubit subspace considered in the main text, $d=2$, and hence
\begin{equation}
    \frac{1}{4} \Phi^{(n,2)} \leq \epsilon^{(2)} \leq \frac{1}{3} \Phi^{(n,2)} .
    \label{eq:sup_infidelity_bounds_qubit}
\end{equation}
Combining Eq.~\eqref{eq:sup_infidelity_bounds_qubit} with Eq.~\eqref{eq:sup_full_and_quadratic_quantities}, the corresponding
relation between the complete quantities can be written as
\begin{equation}
    \frac{1}{4} \Phi^{(n)} \left( \cv_\star^{(n)} \right) \lesssim \epsilon \lesssim \frac{1}{3} \Phi^{(n)} \left( \cv_\star^{(n)} \right),
    \label{eq:sup_infidelity_phi_bounds}
\end{equation}
where $\lesssim$ indicates that the bounds hold to leading quadratic order in the optimized residual generator.  Beyond quadratic
order, the average fidelity error and the compressed Magnus functional generally contain different fourth- and higher-order
invariants of the residual generator.  In particular, dynamics internal to the complementary subspace can enter the fidelity error
through higher-order excursions from the subspace of interest into its complement and back.  Additional corrections are contained
in $\delta\epsilon_{>n}$ and arise from terms beyond the retained Magnus order.

The bounds in Eq.~\eqref{eq:sup_infidelity_phi_bounds} are therefore local to the perturbative regime, where the quadratic
contributions dominate.  In this regime, the compressed Magnus functional and the leading average fidelity error are equivalent
positive quadratic measures of the residual error: each bounds the other by dimension-dependent constants.  Minimizing
$\Phi^{(n)}$ consequently controls the average fidelity error while retaining the generator-level perturbative structure of the
Magnus construction.

\section{Finite-bandwidth filtering of modified pulses}
\label{app:filtering}

This section describes the filtering analysis used to assess the sensitivity of the fully parameterized and compressed pulses to
the finite-bandwidth model considered in the main text. The Gaussian filter used here is an illustrative bandwidth model rather
than a measured transfer function of a specific experimental control chain. It captures the generic attenuation of high-frequency
baseband components but does not include hardware-specific effects such as an asymmetric phase response, ringing, impedance
mismatch, mixer nonlinearity, waveform sampling, or frequency-dependent group delay. The filtering results should therefore be
interpreted as a comparison of the relative spectral sensitivity of the two pulse families, rather than as a prediction of their
performance on a particular experimental platform.

The total modified microwave pulse is written in terms of its two quadratures,
\begin{equation}
    f_{\mm{mod}} (t) = f_{x,\mm{mod}}(t) \cos(\omega_\ud t) + f_{y,\mm{mod}}(t)\sin(\omega_\ud t), 
    \label{eq:app_modified_pulse_quadratures}
\end{equation}
where $f_{j,\mm{mod}} (t)$, with $j=x,y$, includes both the original uncorrected envelope and the Magnus correction, i.e.,
\begin{equation}
    \begin{aligned}
        f_{x,\mm{mod}} (t) &= f_x (t) + g_x (t),\\
        f_{y,\mm{mod}} (t) &= f_y (t) + g_y (t).
    \end{aligned}
    \label{eq:app_mod_env}
\end{equation}

To model finite control bandwidth, we filter the complete modified baseband envelopes $f_{x,\mm{mod}}(t)$ and $f_{y,\mm{mod}}(t)$.
Thus the filter acts on the slowly varying in-phase and quadrature envelopes after the corrections $g_x(t)$ and $g_y(t)$ have been
added; it is not applied to the oscillating lab-frame carrier at $\omega_\ud$. The carrier frequency, including the constant
detuning used to define it, is left unchanged.

We define each modified envelope on $0 \leq t \leq t_\uf$ and extend it by zero outside this interval. The envelopes used here
vanish at $t=0$ and $t=t_\uf$, so this extension introduces no discontinuity in their amplitudes at the boundaries. We then
evaluate the continuous Fourier transform
\begin{equation} 
    \tilde{f}_{j,\mm{mod}}(\omega) = \int_{-\infty}^{\infty} \di{t} f_{j,\mm{mod}}(t) \exp(-i \omega t), \quad j=x,y. 
    \label{eq:sup_modified_pulse_fourier} 
\end{equation} 
Because the filtering is formulated using a continuous Fourier transform rather than a discrete Fourier transform, no numerical
zero-padding or periodic continuation of the pulse is introduced.

The finite-bandwidth response is modeled by the Gaussian transfer function
\begin{equation} 
    F_{\mm{BW}}(\omega) = \exp\left[ -\frac{\log(2)}{2} \left( \frac{\omega}{\omega_{\mm{BW}}} \right)^2 \right].
    \label{eq:sup_gaussian_filter} 
\end{equation}
The prefactor $\log(2)/2$ fixes the bandwidth convention. At $\omega=\omega_{\mm{BW}}$, the amplitude response is
$F_{\mm{BW}}(\omega_{\mm{BW}})=1/\sqrt{2}$ and the corresponding power response is $|F_{\mm{BW}}(\omega_{\mm{BW}})|^2=1/2$. Thus,
$\omega_{\mm{BW}}$ is the $3\,\mathrm{dB}$, or half-power, bandwidth.

The filtered Fourier components are
\begin{equation} 
    \tilde{f}_{j,\mm{mod}}^{\mm{filt}}(\omega) = F_{\mm{BW}}(\omega) \widetilde{f}_{j,\mm{mod}}(\omega), \quad j=x,y, 
    \label{eq:sup_filtered_fourier} 
\end{equation}
and the filtered time-domain envelopes are obtained from 
\begin{equation} 
    f_{j,\mm{mod}}^{\mm{filt}}(t) = \frac{1}{2\pi} \int_{-\infty}^{\infty} \di{\omega} \tilde{f}_{j,\mm{mod}}^{\mm{filt}}(\omega) \exp(i\omega t), \qquad j=x,y. 
    \label{eq:sup_filtered_inverse_fourier}
\end{equation}
Equivalently, this operation convolves each modified envelope with the Gaussian impulse response in the time domain. The filter
therefore suppresses envelope components with frequencies comparable to or larger than $\omega_{\mm{BW}}$ and smooths rapid
temporal features. The filtered envelopes are used directly:~Their amplitudes and pulse areas are not renormalized after
filtering. Consequently, any attenuation or pulse-area change produced by the filter is included in the reported average fidelity
error.

The detuning $\Delta$ is not filtered. It specifies the constant carrier frequency through $\omega_\ud=\omega_{01}+\Delta$ and is
therefore not a time-dependent baseband waveform. For each value of $\omega_{\mm{BW}}$, only $f_{x,\mm{mod}}(t)$ and
$f_{y,\mm{mod}}(t)$ are replaced by their filtered versions; the values of $\omega_\ud$ and $\Delta$ remain fixed.

Although the Gaussian convolution formally produces temporal tails outside $[0,t_\uf]$, the implemented waveform is restricted to
the nominal gate interval and the evolution is evaluated at $t=t_\uf$. The resulting boundary truncation is included in the
reported fidelity error.

For the fully parameterized pulse in Fig.~\ref{fig:pulses} of the main text, the 17 free parameters consist of one static detuning
and 16 Fourier amplitudes, with four sine and four cosine coefficients for each quadrature. Thus the largest explicit harmonic in
the correction envelopes is $\omega_4 = 2\pi 4/t_\uf$. For $\abs{\alpha_2} t_\uf = 5.74$, this gives $\omega_4/\abs{\alpha_2} =
4.38$, explaining why filtering becomes visible once $\omega_\mm{BW}/\abs{\alpha_2}$ is of order a few.

For each value of $\omega_\mm{BW}$, the ideal modified quadratures in the driven-transmon Hamiltonian are replaced by their
filtered versions,
\begin{equation}
    \begin{aligned}
        f_{x,\mm{mod}} (t) &\to f_{x,\mm{mod}}^\mm{filt} (t),\\
        f_{y,\mm{mod}} (t) &\to f_{y,\mm{mod}}^\mm{filt} (t).
    \end{aligned}
    \label{eq:filtered_replacement}
\end{equation}
The filtered quadratures are then inserted into the same Hamiltonian in the frame rotating at the drive frequency as used for the
unfiltered simulations. The carrier at $\omega_\ud$ has been removed in this frame, and the drive frequency enters through the
static detuning $\Delta$.

We then solve numerically the Schrödinger equation in this rotating frame. At the final time $t_\uf$, the filtered evolution is
projected on the computational subspace and the resulting operator is compared with the target gate. The corresponding average
fidelity error is computed using the same definition as in the main text. Repeating this procedure as a function of
$\omega_\mm{BW}$ gives the bandwidth-sensitivity curves in Fig.~\ref{fig:pulses}(d) of the main text.

In the limit $\omega_\mm{BW} \to \infty$, $F_\mm{BW} (\omega) \to 1$ over the spectral support of the pulse, so the filtered
quadratures approach the ideal modified quadratures and the fidelity error approaches the ideal-theory value.  As $\omega_\mm{BW}$
is reduced, high-frequency components of the pulse are suppressed.  A fully parameterized Magnus pulse, which contains more
Fourier components and sharper waveform features, is therefore expected to degrade more rapidly under filtering than a compressed
pulse dominated by a few smooth components.

\end{document}